\documentclass{article}
\usepackage{spconf,amsmath,graphicx}
\usepackage{multirow}
\usepackage{makecell}
\usepackage{bbding}
\usepackage{amssymb}
\usepackage{xcolor}
\usepackage{subcaption}
\usepackage{bibspacing}
\usepackage{hyperref}
\usepackage{marginnote}
\usepackage{bibspacing}
\usepackage{IEEEtrantools}

\def\eg{\emph{e.g. }}

\title{Towards Automatic Data Augmentation for \\Disordered Speech Recognition}
\ninept
\name{Zengrui Jin$^{1}$${\rm \!}$,${\rm\!}$ Xurong Xie$^{2}$${\rm \!}$,${\rm\!}$ Tianzi Wang$^{1}$${\rm \!}$,${\rm\!}$ Mengzhe Geng$^{1}$${\rm \!}$,${\rm\!}$ Jiajun Deng$^{1}$${\rm \!}$,${\rm\!}$ Guinan Li$^{1}$${\rm \!}$,${\rm\!}$ Shujie Hu$^{1}$${\rm \!}$,${\rm\!}$ Xunying Liu$^{1}$}
\address{
\textit{\{zrjin, twang, mzgeng, jjdeng, gnli, sjhu, xyliu\}@se.cuhk.edu.hk, xurong@iscas.ac.cn}
\\$^{1}$ The Chinese University of Hong Kong, Hong Kong SAR, China \\
	$^{2}$ Institute of Software, Chinese Academy of Sciences, China
}

\begin{document}
\bstctlcite{IEEEexample:BSTcontrol}
\setlength{\bibitemsep}{.2\baselineskip plus .05\baselineskip minus .05\baselineskip}

\maketitle
\begin{abstract}
Automatic recognition of disordered speech remains a highly challenging task to date due to data scarcity. 
This paper presents a reinforcement learning (RL) based on-the-fly data augmentation approach for training state-of-the-art PyChain TDNN and end-to-end Conformer ASR systems on such data. 
The handcrafted temporal and spectral mask operations in the standard SpecAugment method that are task and system dependent, together with additionally introduced minimum and maximum cut-offs of these time-frequency masks, are now automatically learned using an RNN-based policy controller and tightly integrated with ASR system training. 
Experiments on the UASpeech corpus suggest the proposed RL-based data augmentation approach consistently produced performance superior or comparable that obtained using expert or handcrafted SpecAugment policies. 
Our RL auto-augmented PyChain TDNN system produced an overall WER of 28.79\% on the UASpeech test set of 16 dysarthric speakers. 
\end{abstract}

\vspace{-0.25cm}
\section{Introduction}
\vspace{-0.25cm}

Despite the rapid progress of automatic speech recognition (ASR) technologies targeting normal speech, accurate recognition of pathological voice, for example, dysarthric speech, remains a challenging task \cite{christensen2012comparative, christensen2013combining, sehgal2015model, yu2018development, geng2020investigation, hu2022exploiting} due to: a) the scarcity of such data; b) their large mismatch against normal speech; and c) large speaker level diversity.  
The physical disabilities and mobility limitations associated with impaired speakers increase the difficulty of collecting large quantities of disordered speech for ASR system development. As a common form of speech disorder, dysarthria is caused by motor control conditions such as cerebral palsy, amyotrophic lateral sclerosis, stroke and traumatic brain injuries \cite{lanier2010speech}. 
%% Disordered speech presents challenges the current deep learning based ASR systems targeting normal speech recorded from healthy speakers due to a significant mismatch.  Also, due to the physical disabilities and mobility limitations observed among impaired speakers, it is difficult to collect large quantities of their speech required for ASR system development.

To this end, data augmentation techniques play a vital role in addressing the above data sparsity issue.
They have been widely investigated in speech recognition systems targeting normal speech. 
By applying these approaches, for example, tempo, vocal tract length or speed perturbation \cite{verhelst1993overlap, kanda2013elastic, jaitly2013vocal, ko2015audio}, simulation of noisy and reverberated speech to improve environmental robustness \cite{ko2017study}, stochastic feature mapping \cite{cui2015data} and back translation in end-to-end systems \cite{hayashi2018back}, the limited ASR system training data is expanded to enhance their generalization.
Recent researches on SpecAugment-based normal speech augmentation apply handcrafted, stochastically generated, or policy-driven temporal and spectral deformation transformations to the original speech data \cite{Park2019, park2020specaugment, hu2021sapaugment, jain22_interspeech, policy2022rui, song2022trimtail, wang2023g, zaiem2023automatic}.

In contrast, only limited research has been conducted on data augmentation targeting disordered speech recognition tasks. 
To modify the overall spectral shape and speaking rate, signal-based approaches including tempo, speed or vocal tract length perturbation \cite{verhelst1993overlap, kanda2013elastic, jaitly2013vocal, ko2015audio, feifei2019phonetic, geng2020investigation} have been developed.
%% derived to transform normal speech data that are widely available into ``disordered like'' speech. 
To capture more detailed spectra-temporal differences between normal and impaired speech, generative adversarial networks (GANs) based speech augmentation \cite{jin2021adversarial, harvill2021synthesis} approaches were also proposed. 
However, the above disordered speech augmentation approaches suffer from the same limitation. 
Neither the tempo, speed or vocal tract length perturbation factors, nor the GAN-based dysarthric speech augmentation models, were trained consistently with the back-end ASR systems. 
Instead, they are derived separately using either phonetic alignment \cite{verhelst1993overlap, kanda2013elastic, jaitly2013vocal, ko2015audio, feifei2019phonetic, geng2020investigation}, or spectral similarity measures \cite{jin2021adversarial, harvill2021synthesis}, but not optimized using the ASR error rate. 
%% , or expert domain knowledge and empirical selection, for example, when setting the temporal or spectral masking hyper-parameters in SpecAugment. 

To this end, a reinforcement learning (RL) based automatic and on-the-fly disordered speech data augmentation approach is proposed for training state-of-the-art end-to-end ASR systems on such data. 
The handcrafted or empirically set temporal and spectral deformation operations in the conventional SpecAugment method are now learned and predicted using an RNN-based policy controller and tightly integrated with the ASR system training process. 
The resulting RL-based augmentation policies (\eg the numbers and sizes of time/frequency masks, and time warping factors) and additionally introduced minimum and maximum cut-offs of these masks, are dynamically learned during ASR system training, commensurate with its generalization to policy injected data diversity. 
An alternating update procedure is applied. 
The first stage involves the policy sampling and reward generation from the current RNN policy controller throughout each training epoch, and the update of the ASR system parameters. 
In the second stage, the RNN controller is updated using the policy rewards collected in the first stage. 
 
Experiments were conducted on the largest benchmark UASpeech dysarthric speech corpus \cite{kim2008dysarthric} using state-of-the-art LF-MMI PyChain TDNN \cite{shao2020pychain} and end-to-end Conformer ASR \cite{watanabe2018ESPnet} systems. 
The automatically learned on-the-fly data augmentation policies consistently outperform the baseline speed perturbation based data augmentation, handcrafted SpecAugment and randomly selected augmentation policies by word error rate (WER) reductions up to 0.5\% absolute (1.71\% relative). 
The best performing automatically data augmented PyChain TDNN \cite{shao2020pychain} system produced an overall WER of 28.79\% on the UASpeech test set containing 16 dysarthric speakers.  

The major contributions of this paper are as follows. 
First, to the best of our knowledge, this paper presents the first use of reinforcement learning based automatic on-the-fly data augmentation approaches targeting disordered speech recognition. 
In contrast, prior researches on automatic data augmentation were predominantly studied for computer vision tasks \cite{cubuk2019autoaugment, lim2019fast, cubuk2020randaugment, hataya2020faster, zhang2020adversarial}.
Second, a novel set of minimum and maximum time/frequency mask cut-offs are proposed to expand the original SpecAugment operations. 
These serve to enhance the power and modelling granularity of RL-based data augmentation policies.
Third, the genericity of our approaches and the accompanying implementation issues also allow them to be applied to other ASR models and task domains in the future.
Finally, the observation on the automatically learned augmentation policies from dysarthric speech provides further insights for practical design of data augmentation techniques tailored for ASR system development using such data. 

\begin{table}[!t]
	\centering
        \renewcommand\arraystretch{0.9}
	\renewcommand\tabcolsep{1pt}
	\caption{Three default hyper-parameters of SpecAugment (ID 1-3), $m_T$ and $m_F$ represent the number of time and frequency masks, $T$ and $F$ stand for the size of time and frequency masks. $W$ is the time warping factor. To further expand the search space of RL-based policy, four additional masking strategies are added as operation 4-7 exclusively for the RL search space. $\dagger$ stands for the search space for the UASpeech dataset and $\clubsuit$ represents the optimal handcrafted hyper-parameters obtained using exhaustive search for the UASpeech dataset. $\diamond$ and $\star$ stand for the optimal hyper-parameters from \cite{Park2019} for the LibriSpeech \cite{panayotov2015librispeech} and SwitchBoard \cite{godfrey1992switchboard}.}
	\vspace{-0.25cm}
	\resizebox{1\linewidth}{!}{
\begin{tabular}{c|c|c|c}
		\hline
		\hline
		ID & Operation & Param. & Values \\
		\hline
		\multirow{2}{*}{1} & \multirow{2}{*}{Time Msk} & $m_T$ & $\{1^\clubsuit, 2, 3, 4, 5\}^\dagger$, $1^{\diamond}$, $2^{\star}$\\
		\cline{3-4}
 		&	& $T$ & $\{1, 2, 3, 4, 5, 6, 7, 8, 9, 10^\clubsuit\}^\dagger$, $100^{\diamond}$, $70^{\star}$ \\
		\hline
		\multirow{2}{*}{2} & \multirow{2}{*}{Freq. Msk} & $m_F$ & $\{1^\clubsuit, 2, 3, 4, 5\}^\dagger$, $1^{\diamond}$, $2^{\star}$\\
		\cline{3-4}
 		&	& $F$ & $\{1, 2, 3, 4, 5, 6, 7, 8, 9, 10^\clubsuit\}^\dagger$, $27^{\diamond}$, $15^{\star}$\\
            \hline
		3 & Time  Warping & $W$ & $\{10, 15, 20^\clubsuit, 25, 30, 35, 40, 45, 50, 55\}^\dagger$, $80^{\diamond}$. $40^{\star}$\\
 		\hline
            \hline
		\multirow{2}{*}{4} & \multirow{2}{*}{Max Time Msk} & $m_T^{\rm max}$ & $\{1, 2, 3, 4, 5\}^\dagger$ \\
		\cline{3-4}
 		&	& $T^{\rm max}$ & $\{1, 2, 3, 4, 5, 6, 7, 8, 9, 10\}^\dagger$  \\
		\hline
		\multirow{2}{*}{5} & \multirow{2}{*}{Max Freq. Msk} & $m_F^{\rm max}$ & $\{1, 2, 3, 4, 5\}^\dagger$ \\
		\cline{3-4}
 		&	& $F^{\rm max}$ & $\{1, 2, 3, 4, 5, 6, 7, 8, 9, 10\}^\dagger$ \\
 		\hline
		\multirow{2}{*}{6} & \multirow{2}{*}{Min Time Msk} & $m_T^{\rm min}$ & $\{1, 2, 3, 4, 5\}^\dagger$ \\
		\cline{3-4}
 		&	& $T^{\rm min}$ & $\{1, 2, 3, 4, 5, 6, 7, 8, 9, 10\}^\dagger$ \\
		\hline
		\multirow{2}{*}{7} & \multirow{2}{*}{Min Freq. Msk} & $m_F^{\rm min}$ & $\{1, 2, 3, 4, 5\}^\dagger$ \\
		\cline{3-4}
 		&	& $F^{\rm min}$ & $\{1, 2, 3, 4, 5, 6, 7, 8, 9, 10\}^\dagger$ \\
		\hline
		\hline
	\end{tabular}
	}
	\label{tab:spec_augment}
	\vspace{-0.6cm}
\end{table}

\vspace{-0.25cm}
\section{Disordered Speech Data Augmentation}
\vspace{-0.25cm}
\label{sec:speed}
Data augmentation targeting disordered speech has been widely investigated. These are predominantly rooted in signal-based augmentation approaches including tempo-stretching \cite{vachhani2018data, xiong2019phonetic}, speed perturbation \cite{ko2015audio} and vocal track length perturbation \cite{jaitly2013vocal}. Among these, speed perturbation \cite{ko2015audio, geng2020investigation, liu2021recent} has been found to be the most effective data augmentation technique for disordered speech. Speed perturbation modifies the original time domain speech signal $x(t)$ by scaling the sampling resolution using a factor $\alpha$. The resulting speed perturbed signal $y(t)$ is given by $y(t)=x(\alpha t)$.
% \begin{equation}
% \setlength{\abovedisplayskip}{0.1pt}
% \setlength{\belowdisplayskip}{0.1pt}
% \resizebox{0.25\linewidth}{!}{
% \begin{math}
% \begin{aligned}
% 	y(t)=x(\alpha t)
% \end{aligned}
% \end{math}
% }
% \end{equation}
Such time-domain modification is equivalent to the following performing in the frequency domain: $X(f) \rightarrow \frac{1}{\alpha}X(\frac{1}{\alpha}f)$,
% \begin{equation}
% \setlength{\abovedisplayskip}{0.1pt}
% \setlength{\belowdisplayskip}{0.1pt}
% \resizebox{0.3\linewidth}{!}{
% \begin{math}
% \begin{aligned}
% 	X(f) \rightarrow \frac{1}{\alpha}X(\frac{1}{\alpha}f)
% \end{aligned}
% \end{math}
% }
% \end{equation}
where $X(f)$ and $\frac{1}{\alpha}X(\frac{1}{\alpha}f)$ denote the Fourier transform of $x(t)$ and $y(t)$ respectively.
Speed perturbation changes the audio duration and the overall spectral shape to simulate a slower speaking rate, reduced speech volume and changes of formant positions of impaired speakers.

During disordered speech augmentation, a set of speaker independent (SI) speed perturbation factors, $\{0.9, 1.0, 1.1\}$, are applied to expand the limited impaired speech. 
In addition, speaker dependent (SD) perturbation is also applied to normal speech. The speaker level perturbation factors are obtained by analysing the phonetic alignment between parallel normal and impaired speech utterances  \cite{xiong2019phonetic, geng2020investigation}. Following \cite{xiong2019phonetic,  geng2020investigation, liu2021recent, jin2021adversarial}, the combined use of SI and SD speed perturbation serve as the baseline default data augmentation approach of this paper.
\vspace{-0.25cm}
\section{SpecAugment}
\vspace{-0.25cm}

The widely used data augmentation approach of {\bf SpecAugment} \cite{Park2019,park2020specaugment} contains $3$ categories of spectra and temporal deformation operations: 1) Time Masks; 2) Frequency Masks; and 3) Time Warping (ID 1-3 in Tab. \ref{tab:spec_augment}). 
Masking operations use utterance-wise mean values to mask randomly selected frequency components and frames.
The numbers and sizes of time and frequency masks are decided by the hyper-parameters $m_T$, $m_F$ and $T$, $F$, respectively. 
Time warping is applied to input acoustic features, by which $W$ interpolated frames are inserted at a randomly selected time step.

The practical use of SpecAugment requires the above large set of hyper-parameters to be manually tuned on specific task domains. For example, the two different sets of handcrafted optimal SpecAugment hyper-parameters suggested in \cite{Park2019} for the benchmark LibriSpeech \cite{panayotov2015librispeech} and SwitchBoard \cite{godfrey1992switchboard} datasets are marked $\diamond$ and $\star$ in Tab. \ref{tab:spec_augment}, respectively.
A set of handcrafted SpecAugment hyper-parameters obtained using an exhaustive search for the UASpeech dataset are also shown in Tab. \ref{tab:spec_augment} and marked by $\clubsuit$. For each of the hyper-parameter, the respective search range is also indicated using "$\{\}^{\dagger}$". These also serve as the reinforcement learning based automatic data augmentation policy search space of the following Sec. \ref{sec:autoaug}.

\vspace{-0.25cm}
\section{Automatic Data Augmentation}
\vspace{-0.25cm}
\label{sec:autoaug}

\begin{figure*}[!t]
  \centering
  \includegraphics[width=\linewidth]{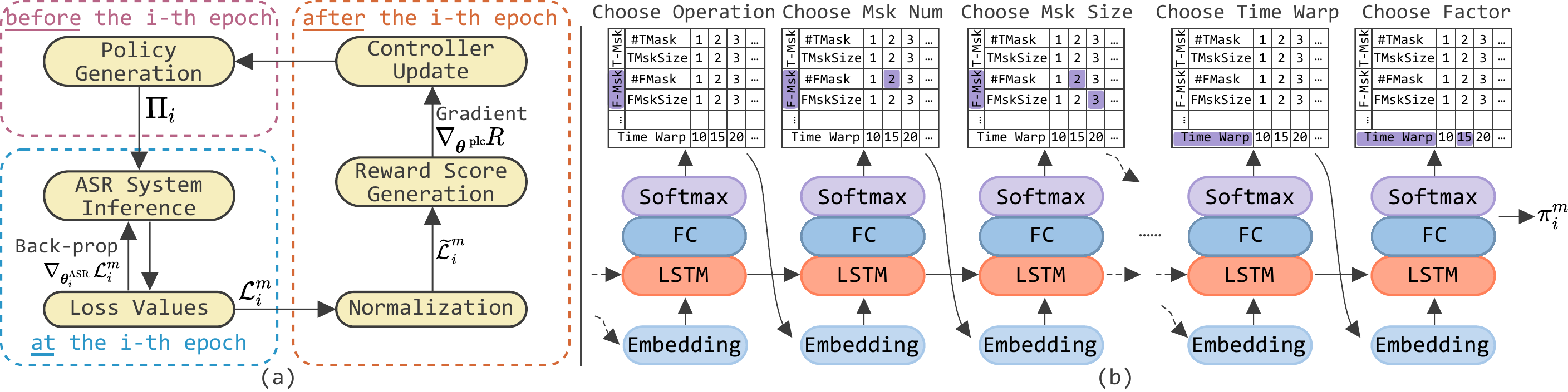}
  \vspace{-0.75cm}
  \caption{Illustration of (a) an overall pipeline for reinforcement learning (RL) based data augmentation policy learning; and (b) the RNN-based RL controller generating a sequence of data augmentation policies of Tab. \ref{tab:spec_augment}. A partial policy sequence constraint is enforced: the sampling always starts with selecting one of seven operations presented in Tab. \ref{tab:spec_augment}. The number of frequency or time masks are sampled immediately afterwards, before sampling their respective sizes.}
  \label{fig:proc}
  \vspace{-0.7cm}
\end{figure*}

In this section, a reinforcement learning (RL) \cite{williams1992simple} based automatic and on-the-fly data augmentation approach is proposed for training ASR systems on disordered speech. 
An alternating update procedure is applied. 
The first stage involves data augmentation policy sampling and reward generation from the current RNN policy controller parameterised by $\boldsymbol{\theta}^{\rm plc}$ within each training epoch, and the update of the ASR system parameters  $\boldsymbol{\theta}^{\rm ASR}$. 
In the second stage, the RNN controller is updated using the policy rewards collected in the first stage. 
Such alternating update procedure is shown in Fig. \ref{fig:proc}(a). 

%% Given the RNN-based controller $\psi$ for policy generation and ASR system $\phi$, $\boldsymbol{\theta}^{\rm plc}$ and $\boldsymbol{\theta}^{\rm ASR}$ represent their parameters respectively.
\noindent
\textbf{Reward Function:}
During the proposed RL-based automatic data augmentation that assists the ASR system training, the expected reward score $R(\cdot)$ at the $i$-th epoch is given by
\vspace{-0.25cm}
\begin{equation}
\resizebox{.85\linewidth}{!}{
\begin{math}
\begin{aligned}
		R(&\boldsymbol{\theta}_{i-1}^{\rm plc},\boldsymbol{\theta}_{i}^{\rm ASR}) \\
  & =\mathbb{E}_{p_{\rm plc} \left(\pi_{i}^m|{\bf \pi}_{(i-1):1}^m;\boldsymbol{\theta}_{i-1}^{\rm plc} \right)}\left[ \mathcal{L}_i^m(\boldsymbol{D}, \pi_{i}^m, \boldsymbol{\theta}_i^{\rm ASR}, \boldsymbol{W}) \right]
\end{aligned} 
\end{math}
}
\label{eq:R}
\vspace{-0.25cm}
\end{equation}
where $\pi_{i}^m$ represents the $m$-th RL searched augmentation policy. 
It is sampled from the RNN controller modelled policy sequence distribution $p_{\rm plc}(\pi_{i}^m | {\bf \pi}_{(i-1):1}^m;\boldsymbol{\theta}^{\rm plc}_{i-1})$ that has been updated during the previous $(i-1)$-th training epoch, and also conditioned on the preceding policy history ${\bf \pi}_{(i-1):1}^m$. 
Each sampled augmentation policy consists of $3$ operations from Tab. \ref{tab:spec_augment}\footnote{An example of $\pi_{i}^m$: [TimeWarp($W$=20), MinTimeMsk($m^{\rm min}_T$=2, $T^{\rm min}$=7), MaxFreqMsk($m^{\rm max}_F$=1, $F^{\rm max}$=3)]} and their associated hyper-parameter settings chosen from the respective search ranges indicated using ``$\{\}^{\dagger}$'' in Tab. \ref{tab:spec_augment}.
%% , and are applied to input acoustic features during the $i$-th training epoch.
%% $p_{\rm plc}(\pi_{i,m} | \pi_{(i-1):1,m};\boldsymbol{\theta}^{\rm plc}_{i-1})$ denotes the RNN controller predicted probability of the sampled policy $\pi_{i,m}$. 
$\mathcal{L}_{i}^m$ denotes the ASR training loss function computed under the sampled augmentation policy $\pi_{i}^m$ for the $i$-th epoch's data. 
$\boldsymbol{D}$ and $\boldsymbol{W}$ stand for the input acoustic features and corresponding word labels respectively.

% \vspace{-0.25cm}
% \subsection{ASR System Update}
% \label{sec:subsec:asr_system_update}
% \vspace{-0.25cm}
%% Configurations for SpecAugment are sampled using the RNN-based controller $\psi$ as shown in Fig. \ref{fig:subfig:controller_network}.
%% The list of tokens that $\psi$ predicts the $i$-th epoch can be used to formulate $M$ SpecAugment configurations $\Pi_i = [\pi_{i,1}, \pi_{i,2}, \cdots, \pi_{i,M}]$, where $M$ represents the number of SpecAugment configurations sampled by $\psi$.
\noindent
\textbf{ASR System Update:}
At the $j$-th minibatch of the $i$-th training epoch, the ASR system parameters are updated via standard back-propagation using gradients that are aggregated over a total of $M$ sampled augmentation policies $\Pi_i = [\pi_{i}^1, \pi_{i}^2, \cdots, \pi_{i}^M]$ and their corresponding perturbed training data. 
This is given by 
\vspace{-0.25cm}
\begin{equation}
\resizebox{.85\linewidth}{!}{
\begin{math}
\boldsymbol{\theta}_{i,j+1}^{\rm ASR} = \boldsymbol{\theta}_{i,j}^{\rm ASR} + \eta^{\rm ASR} \frac{1}{M} \sum\limits^M\limits_{m=1} \nabla_{\boldsymbol{\theta}_{i,j}^{\rm ASR}} \mathcal{L}_{i,j}^m(\boldsymbol{D}, \pi_{i}^m, \boldsymbol{\theta}_{i,j}^{\rm ASR}, \boldsymbol{W}) 
\end{math}
}
\vspace{-0.25cm}
\end{equation}
where $\eta^{\rm ASR}$ is the learning rate for the ASR system update.
%% where $\Delta \boldsymbol{\theta}^{\rm ASR}_{i,m}$ represents the amount of change in $\boldsymbol{\theta}^{\rm ASR}_{i}$ brought by gradient descent on $\pi_{i,m}(\boldsymbol{D})$ of the $i$-th epoch.

\noindent
\textbf{Reward Collection and Controller Update:}
At the end of the $i$-th training epoch, a total of $M$ ASR training loss scores $\{{\cal L}_{i}^1, {\cal L}_{i}^2,\cdots, {\cal L}_{i}^M\}$ are collected for each of the $M$ sampled augmentation policies, where for brevity ${\cal L}_{i}^m = \mathcal{L}_{i}^m(\boldsymbol{D}, \pi_{i}^m, \boldsymbol{\theta}_i^{\rm ASR}, \boldsymbol{W})$ 
is the ASR system training loss score $\mathcal{L}$ associated with the $m$-th sampled policy $\pi_{i}^m$. The RNN controller parameters $\boldsymbol{\theta}^{\rm plc}$ are updated as
\vspace{-0.25cm}
\begin{equation}
\begin{aligned}
\boldsymbol{\theta}_{i}^{\rm plc}=\boldsymbol{\theta}_{i-1}^{\rm plc} - \eta^{\rm plc} \nabla_{\boldsymbol{\theta}^{\rm plc}} R(\boldsymbol{\theta}_{i-1}^{\rm plc}, \boldsymbol{\theta}^{\rm ASR}_i)
\end{aligned}
\vspace{-0.25cm}
\end{equation}
where $\eta^{\rm plc}$ is the learning rate for the controller update. 
The reward gradient with respect to the RNN parameters is calculated by averaging over the sampled $M$ augmentation policies, 
\vspace{-0.25cm}
\begin{equation}
\resizebox{0.85\linewidth}{!}{
\begin{math}
\begin{aligned}
\nabla_{\boldsymbol{\theta}^{\rm plc}} R&(\boldsymbol{\theta}_{i-1}^{\rm plc}, \boldsymbol{{\theta}}^{\rm ASR}_i) 
=\nabla_{\boldsymbol{\theta}^{\rm plc}} \mathbb{E}_{p_{\rm plc}\left( \pi_{i}^m | \pi_{(i-1):1}^m;\boldsymbol{\theta}^{\rm plc}_{i-1}\right)} \left[ {\mathcal{L}}_{i}^m \right] \\
& = \nabla_{\boldsymbol{\theta}^{\rm plc}} \!\! \sum_{\pi_{i}^m}p_{\rm plc}(\pi_{i}^m | \pi_{(i-1):1}^m;\boldsymbol{\theta}^{\rm plc}_{i-1}) \cdot {\mathcal{L}}_{i}^m \\
& \approx \frac{1}{M} \!\!\! \sum_{m=1}^M \!\! {\mathcal{L}}_{i}^m \nabla_{\boldsymbol{\theta}^{\rm plc}}\log p_{\rm plc}(\pi_{i}^m | \pi_{(i-1):1}^m; \boldsymbol{\theta}^{\rm plc}_{i-1}) 
\end{aligned}
\end{math}
}
\vspace{-0.25cm}
\label{eq:reward_score_gradient}
\end{equation}
%% where $\mathcal{L}_{i,m}$ represents the training loss of the ASR system $\mathcal{L}$ associated with the $m$-th SpecAugment configuration $\pi_{i,m}$.
% \begin{equation}
% \mathcal{L}_m = Loss^{\rm ASR}(\mathcal{\phi}(\pi_{i,m}(\boldsymbol{D}), \boldsymbol{\theta}_i^{\rm ASR}), \boldsymbol{W})    
% \end{equation}

\noindent
\textbf{Implementation Issues:}
% {\bf 1) ASR loss score normalization:} To ensure the stable convergence of the RNN controller update, the $M$ ASR training loss scores $\{{\cal L}_{i}^1, {\cal L}_{i}^2,\cdots, {\cal L}_{i}^M\}$ at the $i$-th epoch
% are mean and variance normalized as follows
% \vspace{-0.25cm}
% \begin{equation}
% \begin{aligned}
% 	{\widetilde{{\cal L}}}_{i}^m = \frac{\mathcal{L}_{i}^m - \frac{1}{M}\sum^M_{m=1}\mathcal{L}_{i}^m}{\sqrt{\frac{1}{M}\sum^M_{m=1}(\mathcal{L}_{i}^m - \frac{1}{M}\sum^M_{m=1}\mathcal{L}_{i}^m)^2}}
%  \end{aligned}
%  \vspace{-0.25cm}
% \end{equation}
% before being used in the Eqn. (\ref{eq:reward_score_gradient}) for the RNN controller update.
{\bf 1) ASR loss score normalization:} To ensure the stable convergence of the RNN controller update, the $M$ ASR training loss scores $\{{\cal L}_{i}^1, {\cal L}_{i}^2,\cdots, {\cal L}_{i}^M\}$ at the $i$-th epoch
are mean and variance normalized across $M$ policies before being used in the Eqn. (\ref{eq:reward_score_gradient}) for the RNN controller update.

\noindent
{\bf 2) Number of policy samples:} 
An ablation study on using varying numbers of policy samples during RL-based on-the-fly data augmentation and their performance comparison against using comparable numbers of randomly drawn augmentation policies are presented later in Tab. \ref{tab:ablation}. Empirically the number of policy samples is set as $M=4$ unless otherwise stated. \\
\noindent
{\bf 3) Search space expansion: } To enhance the modelling granularity of data augmentation policies, additional maximum and minimum value based time and frequency masks (row 4-7 in Tab. \ref{tab:spec_augment}) are also applied during time and frequency masking operations. Such masks utilize the utterance level maximum/minimum values instead of the mean of the input features to implement time or frequency masking.

\vspace{-0.25cm}
\section{Experiments}
\vspace{-0.25cm}
All experiments are conducted on the largest publicly available disordered speech dataset UASpeech \cite{kim2008dysarthric}.
It contains 102.7 hours of speech from 16 dysarthric and 13 control speakers with 155 common words and 300 uncommon words for single word recognition.
Utterances are divided into 3 blocks, each containing all common words and one-third of the uncommon words. 
Block 1 and 3 are treated as the training set while block 2 of the 16 dysarthric speakers is treated as the test set. 
Without data augmentation (DA), the training set after silence stripping contains 99195 utterances, around 30.6 hours. 
The test set contains 26520 utterances, around 9 hours.

\begin{table*}[!t]
\centering
\caption{Performance of PyChain TDNN and End-to-end Conformer systems on the 16 UASpeech dysarthric speakers incorporating with different data augmentation strategies. ``Spd.'' represents speed perturbation \cite{geng2020investigation} in Sec. \ref{sec:speed}. ``$W$/$m_F$/$F$/$m_T$/$T$'' represents parameters of the standard SpecAugment operations (Std. SpecAug Op.), ``Ext.'' denotes an expanded set of operations (ID 4-7 in Tab. \ref{tab:spec_augment}). Reported results of Sys. 5-7, 13-16, 20 and 21 are the mean and standard deviation of 5 experiments with different initialization. Number of augmentation policy samples $M = 4$. $\star$ and $\dagger$ denote statistically significant (MAPSSWE \cite{gillick1989some}, $\alpha = 0.05$) WER difference obtained over baselines using speed perturbation only (Sys. 2, 9, 17) and handcrafted SpecAugment (Sys. 4, 12, 19) respectively. For systems using RL-based data augmentation (Sys. 7, 15, 20), the systems producing the median average WERs during 5 random initializations were used in statistical significant tests.}
\vspace{-0.35cm}
\renewcommand\arraystretch{1.}
\resizebox{1.\linewidth}{!}{
\begin{tabular}{l|c|c|c|c|c|c|c|c|c|c|c|c|c}
\hline
\hline
\multirow{3}{*}{Sys.} &
  \multicolumn{3}{c|}{Data Augmentation} &
  \multirow{3}{*}{\# Hrs.} & 
  \multirow{3}{*}{Model} &
  \multirow{3}{*}{\makecell[c]{Spkr. \\ Adapt.}} &
  \multicolumn{7}{c}{Word Error Rate (\%)} \\
   \cline{2-4}
   \cline{8-14}
 & \multirow{2}{*}{Spd.} & SpecAugment & \multirow{2}{*}{Ext.} & & & & \multirow{2}{*}{Very Low} & \multirow{2}{*}{Low} & \multirow{2}{*}{Medium} & \multirow{2}{*}{High} & \multirow{2}{*}{Seen} & \multirow{2}{*}{Unseen} & \multirow{2}{*}{Average} \\
   \cline{3-3}
 & & $W$/$m_F$/$F$/$m_T$/$T$ & &  &  &  &  &  &  &  & & \\ 
   \hline
1 & - & - & - & 30.6 & \multirow{7}{*}{\makecell[c]{PyChain \\ TDNN}} & \multirow{7}{*}{-} & 66.34 & 36.20 & 24.13 & 10.27 & 26.56 & 39.32 & 31.57 \\ 
  \cline{1-5}
  \cline{8-14}
2 & \multirow{6}{*}{\checkmark} & - & \multirow{4}{*}{-} & \multirow{6}{*}{130.1} &  &  & 65.40 & 32.11 & 23.06 & 10.15 & 25.02 & 37.91 & 30.02  \\ 
  \cline{1-1}
  \cline{3-3}
  \cline{8-14}
3 \hfill (expert \cite{Park2019}) &  & 80/1/27/1/100  &  &  &  &  & 69.57 & 33.68 & 24.18 & 10.17 & 28.10 & 36.95 & 31.53 \\
  \cline{1-1}
  \cline{3-3}
  \cline{8-14}
4 \hfill (handcrafted) &  & 20/1/10/1/10 & 
 & & &  & \textbf{64.28}$^{\star}$ & 30.25$^{\star}$ & 22.12 & 10.64 & 25.04 & 35.98$^{\star}$ & 29.29$^{\star}$ \\
  \cline{1-1}
  \cline{3-3}
  \cline{8-14}
5 \hfill (Std. SpecAug Op.) & & RL-based ($M=4$) &  & & &  & 65.94${\pm1.05}$ & 31.09${\pm0.99}$ & 23.06${\pm0.96}$ & 9.95${\pm0.51}$ & 25.02${\pm0.66}$ & 37.34${\pm0.82}$ & 29.81${\pm0.54}$ \\
  \cline{1-1}
  \cline{3-4}
  \cline{8-14}
6 \hfill (Ext.) &  & \multicolumn{2}{c|}{Random ($M=4$)} & & &  & 65.65${\pm1.35}$ & 31.89${\pm1.57}$ & 23.43${\pm1.17}$ & 9.55${\pm0.55}$ & 24.48${\pm1.02}$ & 37.76${\pm1.15}$ & 29.47${\pm0.39}$ \\
  \cline{1-1}
  \cline{3-4}
  \cline{8-14}
7 \hfill (Ext.) &  & \multicolumn{2}{c|}{RL-based ($M=4$)} & &  &  & 64.73${\pm0.54}^{\star}$ & \textbf{29.95${\pm0.32}^{\star}$} & \textbf{21.79${\pm0.56}^{\star}$} & \textbf{9.31${\pm0.18}^{\star\dagger}$} & \textbf{24.29${\pm0.32}^{\star\dagger}$} & \textbf{35.89${\pm0.30}^{\star}$} & \textbf{28.79${\pm0.12}^{\star\dagger}$} \\
  \hline
  \hline
8 & \multirow{9}{*}{\checkmark} & - & - & 130.1 & 
  \multirow{9}{*}{\makecell[c]{ESPnet \\ Conformer}} & \multirow{8}{*}{-} & 73.88 & 53.12 & 49.92 & 42.03  & 23.51 & 99.14 & 53.17 \\ 
  \cline{1-1}
  \cline{3-5}
  \cline{8-14}
9 & & - & \multirow{5}{*}{-} & \multirow{8}{*}{190} &  &  & 64.75 & 40.81 & 32.80 & 8.87 & 18.33 & 57.25 & 33.58 \\ 
  \cline{1-1}
  \cline{3-3}
  \cline{8-14}
10 \hfill (expert \cite{Park2019}) &  & 80/1/27/1/100 &  & & &  & 70.94 & 39.82 & 28.47 & 8.85 & 21.97 & 52.13 & 33.80 \\
  \cline{1-1}
  \cline{3-3}
  \cline{8-14}
11 \hfill (from Sys. 4) & & 20/1/10/1/10 &  &  &  &  & 65.70 & 40.63 & 33.39 & 9.53 & 19.03 & 57.40 & 34.07  \\ 
  \cline{1-1}
  \cline{3-3}
  \cline{8-14}
12 \hfill (handcrafted) & & 30/1/5/1/5 &  &  &  &  & 65.45 & \textbf{38.82}$^{\star}$ & \textbf{29.84}$^{\star}$ & 8.84 & 18.15 & \textbf{55.10}$^{\star}$ & \textbf{32.64}$^{\star}$ \\
  \cline{1-1}
  \cline{3-3}
  \cline{8-14}
13 \hfill (Std. SpecAug Op.) & & RL-based ($M=4$) &  &  &  &  & 65.51${\pm0.80}$ & 41.98${\pm0.30}$ & 33.81${\pm0.87}$ & 11.92${\pm0.62}$ & 17.96${\pm0.11}$ & 62.12${\pm0.52}$ & 35.28${\pm0.13}$ \\
  \cline{1-1}
  \cline{3-4}
  \cline{8-14}
14 \hfill (Ext.) & & \multicolumn{2}{c|}{Random ($M=4$)} &  &  &  & 65.56${\pm0.26}$ & 39.09${\pm0.16}$ & 31.47${\pm0.33}$ & \textbf{8.66${\pm0.37}$} & 17.74${\pm0.13}$ & 56.64${\pm0.37}$ & 33.02${\pm0.04}$ \\
  \cline{1-1}
  \cline{3-4}
  \cline{8-14}
15 \hfill (Ext.) & & \multicolumn{2}{c|}{RL-based ($M=4$)}  &  &  &  & \textbf{64.66${\pm0.39}$} & 39.40${\pm0.56}^{\star}$ & 30.19${\pm0.73}^{\star}$ & 8.89${\pm0.00}$ & \textbf{16.93${\pm0.45}^{\star\dagger}$} & 57.25${\pm0.30}$ & 32.68${\pm0.40}^{\star}$ \\  
  \cline{1-1}
  \cline{3-4}
  \cline{7-14}
16 \hfill (Ext.) &  & \multicolumn{2}{c|}{RL-based ($M=4$)}  &  &  & \checkmark & 65.57 & \textbf{37.50}$^{\star\dagger}$ & \textbf{27.78}$^{\star\dagger}$ & 9.54 & 22.07 & \textbf{47.82}$^{\star\dagger}$  & \textbf{32.16}$^{\star\dagger}$ \\  
  \hline
  \hline
17 & \multirow{4}{*}{\checkmark} & - & \multirow{3}{*}{-} &  \multirow{5}{*}{173}  &  \multirow{5}{*}{\makecell[c]{ESPnet \\ Conformer}} &  \multirow{4}{*}{-} & 66.75 & 38.40 & 29.29 & 7.27 & 20.21 & 50.71 & 32.17 \\ 
  \cline{1-1}
  \cline{3-3}
  \cline{8-14}
18 \hfill (from Sys. 4) & \multirow{4}{*}{\checkmark} & 20/1/10/1/10 &  &  &  &  & 66.20 & 39.08 & 28.17 & 6.39 & 20.39 & 49.29 & 31.72 \\ 
  \cline{1-1}
  \cline{3-3}
  \cline{8-14}
19 \hfill (handcrafted) & & 30/1/5/1/5 &  &  &  &  & \textbf{65.36} & 36.99 & 26.37 & 6.70 & 19.11 & 48.81 & 30.76 \\
  \cline{1-1}
  \cline{3-4}
  \cline{8-14}
20 \hfill (Ext.) & & \multicolumn{2}{c|}{RL-based ($M=4$)}  &  &  &  & 66.57${\pm0.15}$ & \textbf{35.97}${\pm1.09}^{\star\dagger}$ & \textbf{26.06}${\pm0.54}^{\star}$ & \textbf{5.45}${\pm0.51}$ & \textbf{19.02}${\pm0.40}^{\star}$ & \textbf{47.71}${\pm1.04}^{\star\dagger}$ & \textbf{30.27}${\pm0.43}^{\star\dagger}$ \\
  \cline{1-1}
  \cline{3-4}
  \cline{7-14}
21 \hfill (Ext.) & & \multicolumn{2}{c|}{RL-based ($M=4$)}  &  &  & \checkmark & 66.66 & \textbf{35.35}$^{\star\dagger}$ & \textbf{25.72}$^{\star}$ & \textbf{5.16}$^{\star\dagger}$ & \textbf{18.98}$^{\star}$ & \textbf{47.00}$^{\star\dagger}$ & \textbf{29.97}$^{\star\dagger}$ \\
  \hline
  \hline
\end{tabular}
}
\label{tab:performance}
\vspace{-0.6cm}
\end{table*}

\noindent
\textbf{Experiment Setup and Baseline Systems:}
The RNN-based controller is implemented as a one-layer LSTM with its hidden size set to $128$ and embedding size set to $32$. 
Adam optimizer is applied with the learning rate set to $0.00035$. 
An entropy penalty with a weight of $0.00001$ is applied to avoid unexpected rapid convergence. 

The LF-MMI factored time delay neural network (TDNN) systems are implemented to predict biphone output units using PyChain ~\cite{shao2020pychain}.
The End-to-end (E2E) Conformer systems\footnote{12 Conformer encoder blocks + 12 Transformer decoder blocks, feed-forward layer dim = 2048, attention heads = 4, attention head dim = 256, conv kernel size = 7} are implemented using the ESPnet toolkit \cite{watanabe2018ESPnet} to directly model letter sequence outputs. 
The HTK toolkit \cite{young2002htk} is used for phonetic alignment in speed perturbation factors estimation \cite{geng2020investigation}, silence stripping and feature extraction. 
Speed perturbation is implemented using SoX.

The SpecAugment hyper-parameters of handcrafted systems (Sys. 4, 12, 19) are obtained using grid search. 
There are 25000 combinations of SpecAugment hyper-parameters in total as suggested in Tab. \ref{tab:spec_augment}.
Exhaustive search over these  is infeasible. Hence, a sequential, two-stage grid search was conducted instead. In the first stage, the optimal $T$/$F$ mask numbers and $T$/$F$ mask sizes were manually searched with the time warping factor set to 1.0 (turned off). To reduce the search cost, the $T$/$F$ mask numbers are further set to be equal, as are the $T$/$F$ mask sizes.
In the second stage, the time warping factor was tuned. This resulted in 25 hyper-parameter combinations explicitly evaluated on Conformer systems.

\noindent
\textbf{Experiments on the PyChain TDNN Systems:}
The performance of various PyChain TDNN systems is shown in Tab. \ref{tab:performance} (Sys. 1-7). 
The following trends can be found in Tab. \ref{tab:performance}:
\textbf{(1)} Directly applying the SpecAugment hyper-parameters handcrafted on LibriSpeech data to UASpeech leads to large WER increases of up to 1.51\% absolute (5.03\% relative) (Sys. 3 \textit{vs.} 2, last col.). 
\textbf{(2)} In addition, SpecAugment hyper-parameters obtained using exhaustive grid search on the UASpeech dataset outperforms the one with speed perturbation only by 0.73\% absolute (2.43\% relative) WER reduction (Sys. 4 \textit{vs.} 2, last col.), indicating the SpecAugment hyper-parameters for TDNN Systems are domain and task dependent.
\textbf{(3)} When the proposed RL-based policy search being applied to the original SpecAugment operation set, it creates a 1.72\% absolute (5.46\% relative) WER reduction against the expert system (Sys. 5 \textit{vs.} 3, last col.).
\textbf{(4)} The proposed RL-based data augmentation outperform both the handcrafted SpecAugment policy (Sys. 7 \textit{vs.} 4, last col.) and the randomly sampled policies (Sys. 7 \textit{vs.} 6, last col.) by 0.5\% and 0.68\% absolute (1.71\% and 2.31\% relative) in terms of overall WER reduction, respectively; 
\textbf{(5)} The performance of the best performing TDNN system using RL-based data augmentation (Sys. 6, Tab. \ref{tab:performance}) is further contrasted with recently published results on the UASpeech task in Tab. \ref{tab:comparison}.

\begin{table}[!t]
    \centering
    \renewcommand\tabcolsep{1pt}
    \renewcommand\arraystretch{1.2}
    \caption{Ablation study on the PyChain TDNN system investigating the impact of the number of RL learned augmentation policy samples $M$ of Sec. \ref{sec:autoaug}, where $M \in \{2, 4, 8\}$.  ``Rd.'' and ``RL'' represents randomly selected, or RL-based Augmentation policies respectively.}
    \vspace{-0.3cm}
    \resizebox{1\columnwidth}{!}{
    \begin{tabular}{c|c|c|c|c|c|c|c|c}
		\hline
		\hline
		\multirow{2}{*}{$M$} & \multirow{2}{*}{Plc.} &  \multicolumn{7}{c}{Word Error Rate (\%)} \\
		\cline{3-9}
		&  & Very Low & Low & Medium & High & Seen & Unseen & Average \\
		\hline
		\multirow{2}{*}{2} & Rd. & 64.71${_\pm 0.43}$ & 31.43${_\pm 0.70}$ & 23.22${_\pm 0.87}$ & 9.83${_\pm 0.56}$ & 24.83${_\pm 1.08}$ & 37.16${_\pm 1.33}$ & 29.62${_\pm 0.36}$ \\
		% \cline{1-1}
		% \cline{2-9}
		& RL & 65.88${_\pm 0.58}$ & 30.93${_\pm 1.54}$ & 22.36${_\pm 0.32}$ & 9.46${_\pm 0.66}$ & 24.81${_\pm 0.34}$ & 36.76${_\pm 0.81}$ & 29.45${_\pm0.34}$ \\
		\hline		
		\multirow{2}{*}{4} & Rd. &65.65${_\pm1.35}$ & 31.89${_\pm1.57}$ & 23.43${_\pm1.17}$ & 9.55${_\pm0.55}$ & 24.48${_\pm1.02}$ & 37.76${_\pm1.15}$ & 29.47${_\pm0.39}$  \\
		% \cline{1-1}
		% \cline{2-9}
		& \textbf{RL} & \textbf{64.73${_\pm0.54}$} & \textbf{29.95${_\pm0.32}$} & \textbf{21.79${_\pm0.56}$} & \textbf{9.31${_\pm0.18}$} & \textbf{24.29${_\pm0.32}$} & \textbf{35.89${_\pm0.30}$} & \textbf{28.79${_\pm0.12}$} \\
		\hline
		 \multirow{2}{*}{8} & Rd. & 65.82${_\pm1.17}$ & 30.35${_\pm1.34}$ & 22.63${_\pm0.71}$ & 9.76${_\pm0.80}$ & 24.68${_\pm1.14}$ & 36.93${_\pm0.86}$ & 29.44${_\pm0.47}$ \\
		 & RL & 65.28${_\pm0.81}$ & 31.36${_\pm0.56}$ & 23.48${_\pm0.70}$ & 10.03${_\pm0.54}$ & 25.02${_\pm0.76}$ & 37.66${_\pm1.30}$ & 29.84${_\pm0.46}$ \\
		\hline
		\hline
	\end{tabular}
	}
	\label{tab:ablation}
	\vspace{-0.6cm}
\end{table}

\noindent
\textbf{Experiments on the End-to-end Conformer Systems:}
As E2E ASR systems are sensitive to the training data coverage (Sys. 9 \textit{vs.} 8 in Tab. \ref{tab:performance}), B2 data of the 13 control speakers are also used in Conformer training. 
This produces a 190-hr augmented training set after applying speaker independent and dependent speed perturbation \cite{geng2020investigation} (173-hr after silence stripping further applied). 
Results of the ESPnet \cite{watanabe2018ESPnet} E2E Conformer systems are presented in Tab. \ref{tab:performance} (Sys. 9 - 16 for the 190-hr training set, Sys. 17 - 21 for the 173-hr training set).
Several trends can be found in Tab. \ref{tab:performance} (Sys. 8-21):
\textbf{(1)} The hyper-parameters of SpecAugment are highly domain and system specific. 
Compared with the baseline Conformer system without SpecAugment (Sys. 9), the optimal SpecAugment hyper-parameters tuned on the LibriSpeech dataset from \cite{Park2019} (Sys. 10) and the hyper-parameters handcrafted on TDNN models (Sys. 11 \& 18, settings from Sys. 4) both lead to performance degradation, while re-conducting exhaustive search on the UASpeech dataset for the Conformer system produces an average WER reductions of {\bf 0.45\%-0.94\%} over the baseline Conformer (Sys. 12 \textit{vs.} 9, Sys. 18 \textit{vs.} 17).
\textbf{(2)} The RL-based automatic data augmentation produces WERs comparable to using handcrafted SpecAugment hyper-parameters (Sys. 15 \textit{vs.} 12, Sys. 20 \textit{vs.} 19), the expanded set of operations also brings an average WER reduction of \textbf{2.6\%} (Sys. 15 \textit{vs.} 13).
\textbf{(3)} The Conformer systems using RL-based data augmentation produce consistent WER reductions over those obtained by expert knowledge or hand-crafting (Sys. 15 \textit{vs.} 9-13, Sys. 20 \textit{vs.} 17-19) on the most challenging ``very low'' intelligibility subgroup by up to 0.79\% absolute (1.21\% relative) WER reduction. 
It also outperforms randomly selected policies by an average WER reduction of 0.34\% absolute (Sys. 15 \textit{vs.} 14).
\textbf{(4)} Further improvements are obtained after applying SBE speaker adaptation \cite{geng2022spectro} to the two Conformer models using RL-based data augmentation (Sys. 16 \textit{vs.} 15, Sys. 21 \textit{vs.} 20).

\noindent
\textbf{Ablation Study:}
An ablation study is conducted to analyze the impact of the number of augmentation policy samples, $M$.
The results presented in Tab. \ref{tab:ablation} suggest that for the TDNN systems, the RL learned augmentation policies give the best performance on all intelligibility when $M = 4$ (Row 4, Tab. \ref{tab:ablation}).
It is also observed that the percentages of operations such as Time Warp, Min Time Mask and Frequency Mask gradually increase along with the number epochs, implying relatively higher importance of these augmentation operations towards the latter phase of ASR system training.

% \begin{figure}[!t]
%   \centering
%   \includegraphics[width=0.97\linewidth]{figs/visualization/vis_crop.pdf}
%   \vspace{-0.25cm}
%   \caption{Percentages of SpecAugment operations over PyChain TDNN system training epochs that are: (a) randomly selected; or (b) RL learned, with the number of policy samples $M=4$ on the UASpeech data. The percentage of each operation is average over five different experiments.}
%   \vspace{-0.75cm}
%    \label{fig:policy}
% \end{figure}

\begin{table}[!h]
    \centering
    \vspace{-2mm}
    \caption{A comparison between published systems on UASpeech and our system. ``L'', ``VL'' and ``Avg.'' represent WER (\%) for low, very low intelligibility subgroups and the average WER.}
    \label{tab:comparison}
    \vspace{-0.35cm}
    \renewcommand\tabcolsep{1pt}
    \resizebox{\columnwidth}{!}{
    \begin{tabular}{c|c|c|c}
    \hline
    \hline
    Systems & VL & L & Avg. \\
    \hline
    CUHK-2018 DNN System Combination \cite{yu2018development} & - & - & 30.60 \\
Sheffield-2019 Kaldi TDNN + DA \cite{xiong2019phonetic} & 67.83 & 27.55 & 30.01 \\
\makecell{Sheffield-2020 CNN-TDNN + speaker adaptation \cite{xiong2020source}} & 68.24 & 33.15 & 30.76 \\
CUHK-2020 DNN + DA + LHUC-SAT \cite{geng2020investigation} & 62.44 & 27.55 & 26.37 \\
CUHK-2021 QuartzNet + CTC + Meta Learning + SAT \cite{wang2021improved} & 69.30 & 33.70 & 30.50 \\
CUHK-2021 DNN + DCGAN + LHUC-SAT \cite{jin2021adversarial} & 61.42 & 27.37 & 25.89 \\
Nagoya Univ.-2022 WavLM \cite{lester2022investigatiing} & 71.50 & 50.00 & 51.80 \\
CUHK-2023 Kaldi TDNN + VAE-GAN + LHUC-SAT \cite{jin2022adversarial} & 57.31 & 28.53 & 27.78 \\
JHU-2023 DuTa-VC (Diffusion) + Conformer \cite{wang23qa_interspeech} & 63.70 & 27.70 & 27.90\\
\hline
\makecell[c]{\textbf{PyChain TDNN + RL-based Augmentation} (Sys. 7, Tab. \ref{tab:performance}) } & 64.73 & 29.95 & 28.79 \\
\hline
\makecell[c]{\textbf{Conformer + RL-based Augmentation} (Sys. 21, Tab. \ref{tab:performance}) } & 66.66 & 35.35 & 29.97 \\
\hline
\hline
\end{tabular}
}
\vspace{-4mm}
\end{table}

\vspace{-0.25cm}
\section{Conclusions}
\vspace{-0.25cm}

This paper proposed a reinforcement learning based on-the-fly data augmentation approach for disordered speech recognition. 
Experiments conducted on the UASpeech suggest the hyper-parameters of the conventional SpecAugment approach are highly domain and system specific. 
The proposed RL-based automatic on-the-fly data augmentation approach alleviates the need of expensive handcrafting when being applied to new speech task domains. 
%% Improved model generalization is also obtained over the baseline speed perturbation and handcrafted SpecAugment policies. 
Future research will improve the interpretability and controllability of RL learned data augmentation policies for personalized data augmentation. 

\newpage
\bibliographystyle{IEEEtran}
\bibliography{mybib,bib/intro,bib/specaug,bib/autoaug}

\end{document}